\documentclass{article}[15pt] 

\usepackage{graphicx,amscd,amsmath,amssymb,verbatim}
\usepackage[dvips]{hyperref}
\usepackage[TS1,OT1,T1]{fontenc}

\begin{document}

\title{Can Born Infeld Gravity Explain  Galaxy  Rotation  Curves? }
\author{Musongela  Lubo \\
D\'epartement de  Physique \\  Facult\'e  des  Sciences \\
Universit\'e de Kinshasa \\
BP 190, Kinshasa  XI \\
R\'epublique D\'emocratique du Congo.\\
musofree@yahoo.com}
\date{\today}

\maketitle
\begin{abstract}

A  Born-Infeld  theory has been proposed in order to explain both dark matter and  dark energy. We show that the  approximation used  to  deduce the flat  rotation  curves in this model  breaks down at  large distances in a spherically symmetric configuration.
\end{abstract}

\section{Introduction}

\noindent  Dark matter and dark  energy  are among the most exciting  topics for  cosmology  and  astrophysics. The first  was introduced  in order to account for the fact that the velocities of  particles and bodies  in galaxies do not fall  with distance as they should  if the only  matter source  present  was visible.  Dark energy on the other side can provide the necessary ingredient to  the understanding of  the late time acceleration of the universe.

\noindent It has been  pointed out that  actually,  in  both  cases  the discrepancy  between   observations and  theory  could  be the signal  that  something is missing in  our  knowledge  of  gravity \cite{{mil1,mil2,bek1,bek2,moffat,jacobson}} .  This path is philosophically different from  the  current  trend which favors new  particle physics  beyond the standard model \cite{{cold1, cold2,cold3,cold4,cold5}} . It has led to different  proposals, each with its own  successes and shortcomings  \cite{{turner,dolgov,mersini,lubo}}  .   

\noindent The interest  of  the  proposal  \cite{{bana}}    is  that, if  correct  it  solves  both  problems  simultaneously.  One can  question  the minimality and motivation   of the  model.  In  the  popular  trend (Lambda CDM  cosmology), one  uses   fields  very  similar  to  ordinary  matter  but  displaying  very  small  interactions with it.  It is especially appealing  to  begin with  supersymmetry  and obtain as a by product  the explanation for the rotation curves of galaxies.  The model  proposed  in  \cite{{bana}}  introduces a new symmetric  second rank tensor. It  does not couple directly to ordinary matter but interacts  with the metric, just like dark matter. 
A more profound  motivation was  sketched.

\noindent In this note  we  analyze    the first  order  approximation  used  in    \cite{{bana}}
  which  led  to  the  claim  that  this  theory  could  explain  the flat  rotation  curves  observed  in  galaxies. We essentially  point  out    that  the approximation  generically  breaks  down far  from  the  source, because  the  first  non vanishing contribution of the expansion  is  bigger  than the  zeroth one in that  region.  One thus needs a more  careful analysis  of the problem, which is not done in this short  letter.

\section{The model}

One considers  the universe to be governed  by  the following  Lagrangian

\begin{equation}
 S =  \frac{1}{16 \pi G}  \int  d^4 x  
\left[
\sqrt{ \vert g \vert}   R +   \frac{2}{\alpha l^2}   \,  \sqrt{    \vert \,   g_{\mu \nu}  \, -    \,  l^2  \, K_{\mu \nu}  \,    \vert  } +
L^{matter} (\psi,g)  \right]    \quad .
\end{equation}

The Riemannian metric is denoted as usual by $g_{\mu \nu}$; $G$ is the Newton constant. There is a new length scale $l$ and a dimensionless constant $ \alpha  $ in the theory. The tensor $  K_{\mu     \nu}  $ is the curvature  of a new connection $ C^{\mu}_{\nu \rho} $. This connection is the Christoffel symbol of  a  new symmetric  second rank tensor   which is denoted $q_{ \alpha  \beta}$ while its inverse is written   $q^{\alpha   \beta}$.

Notice that ordinary, baryonic matter couples directly only to gravity. Its link to the new field is an indirect one. The equations of motion in the gravity sector are

\begin{equation}
\label{first}
G_{\mu  \nu}  =  - \frac{1}{l^2}   \sqrt{ \frac{q}{g} }      g_{\mu  \alpha }  q^{\alpha    \beta} g_{\beta  \nu} + 8 \pi G T^{m}_{\mu  \nu}           \    
\quad {\rm and }  \quad  K_{\mu \nu} =  \frac{1}{l^2} (g_{\mu \nu} + \alpha q_{\mu \nu} )  \quad .  
\end{equation}

 In the absence of ordinary matter, one can find solutions for which there is a constant factor between the two symetric tensor fields. The preceding equations then formally look like usual Einstein gravity with a cosmological constant. De Sitter space time is therefore a solution and this model can explain the late time acceleration of  a Robertson-Walker space time.

\section{The galactic rotation curves}

For a spherically symmetric configuration, the approach taken in    \cite{{bana}}   is much more involved. The  metric and its companion tensor  are  taken to be of the form

\begin{eqnarray}
\label{sixth}
ds^2  &=& - c^2  \left( 1+ \frac{1}{c^2}  \Phi(r)   \right) dt^2 + \left( 1- \frac{2 m(r)}{c^2 r}  \right)^{-1} dr^2
+ r^2 d \Omega^2   ,    \nonumber \\
  q_{\mu \nu} dx^\mu dx^\nu  &=&   - \beta^2 c^2   \left(  1   -  \frac{ \omega_0 }{\tilde{k}(r)}     \right) dt^2 +
   \left(  1   -  \frac{ \omega_0 }{\tilde{k}(r)}     \right)^{-1}       ( {\tilde{k}^{'}}(r))^2            dr^2  \nonumber \\
  & + &    ( {\tilde{k}}(r))^2    d\Omega^2  \quad .
\end{eqnarray}

    The expansion is made in terms of the new length scale:
\begin{eqnarray}
\label{third}
\Phi(r) &=& \Phi^{(0)}(r) + \frac{1}{l^2} \Phi^{(1)}(r) + \frac{1}{l^4} \Phi^{(2)}(r) +  \cdots   \quad ,   \nonumber \\
m(r)    &=& m^{(0)}(r) + \frac{1}{l^2}  m^{(1)}(r) + \frac{1}{l^4} m^{(2)}(r) +  \cdots  \quad .
\end{eqnarray}

The approximation found in \cite{{bana}}   was obtained by  the following procedure. One first excludes all  normal matter since the new field is supposed to play the role of dark energy, which dominates over ordinary matter. Thus the baryonic energy momentum tensor in the first part of Eq(\ref{first}) is taken to be vanishing. Secondly, one considers that the length $l$ being of  cosmological scale, the right hand side  of the second part of   Eq(\ref{first})  can be neglected in the first approximation when dealing with a galaxy. The metric can then be taken to be the Schwarzchild solution with an arbitrary mass, but the vanishing mass was  chosen instead. This  means one has
\begin{equation}
 \Phi^{(0)}(r)    =  m^{(0)}(r)    =  0   \quad .
\end{equation}

The form given to the second tensor is a solution to the field equations  given in    Eq(\ref{first}), in the limit $l \rightarrow \infty$. It works  for arbitrary values of the parameters  $\omega_0 , \beta$ and for an arbitrary function $  {\tilde{k}}(r)$, which is not specified at this point. We will come back to this later.  One then writes the   equations linking the unknown functions  $\Phi^{(1)}(r) , m^{(1)}(r)   $  and $ {\tilde{k}}(r)$ which till now is arbitrary. Making a change of variables, one obtains the following expressions:
\begin{eqnarray}
\label{second}
 r(k)  & =&  A_0    \left( -  \left( k -  \frac{1}{2} \right)      \ln{\left( 1 - \frac{1}{k} \right) } - 1  \right)  + B_0   \left( k -  \frac{1}{2} \right) 
 \quad ,   \nonumber  \\
  u^{(1)}(k)  & =&      \frac{1}{2} \beta c^2  B_0          \left[ A_0   \left(   k^2   \left( 1 -  \frac{1}{k} \right)  \ln{\left( 1 - \frac{1}{k} \right) } +    k -  \frac{1}{2} \right)  - 
B_0 (k^2-k)  \right]  \quad ,   \nonumber  \\
 m^{(1)}(k)  & =&  \frac{\omega_0^3 c^2 }{2 \beta}  \left(   \frac{1}{3}  k^3 + \frac{1}{2}  k^2 +  \ln{(  k - 1 )    }        -  h_0  \right) 
\quad ,
\end{eqnarray}
where the function $u^{(1)}(r ) $  has been defined by the relation
\begin{equation}
\label{fourth}
r   \frac{d \Phi^{(1)}(r)}{dr}  = u^{(1)}(r )  +  \frac{m^{(1)}(r) }{r}   \quad .
\end{equation}

According to the values of the parameters $A_0 , B_0 , \omega_0 $ , the solution can display different behaviors. We shall in this paper confine ourselves to the so called "logarithmic branch" for which one has $A_0 > 0, B_0  < 0$. Under these conditions,  the function $r(k)$ evolves between two extreme values: a  finite value $k_0$ where it vanishes and the value $k=1$    where it diverges.

Looking at  Eq(\ref{second}) , one sees  that the radial coordinate  $r$ goes to infinity as $k $  goes to one. We now  want to derive an approximation  of the different components of the metric in terms of the radius in the  asymptotic region.  Near $k =1$, one can replace $k$ by unity everywhere, except in the term containing the logarithm since  it blows up. Inverting the relation, one  readily finds
\begin{equation}
{\tilde{k}(r)}  = \omega_0   \left[  1 -      \exp{     \left(  - 2 + \frac{B_0}{A_0} \right)    \exp{     \left(  -  \frac{2}{A_0}  r  \right) }     } 
 \right]^{-1}   \quad .
\end{equation}
Using the same argument, one obtains for the other functions the following expressions:
\begin{eqnarray}
\label{five}
 u^{(1)}(r )    &=&   \frac{1}{2} \beta c^2 \omega_0   A_0  \left\lbrace  \exp{     \left(  - 2 + \frac{B_0}{A_0} \right)   
 \exp{     \left(  -  \frac{2}{A_0}  r  \right) }     }  \frac{2}{A_0} 
 \left(  -  r  - A_0 + \frac{1}{2}   B_0  \right) +  \frac{1}{2}   \right\rbrace   \quad ,    \nonumber \\
m^{(1)}(r ) &=&    \frac{ \omega_0^3 c^2}{2 \beta}    \left[   \frac{11}{6} -   h_0   +  \frac{2}{A_0} 
 \left(  -  r  - A_0 + \frac{1}{2}   B_0  \right)    \right]  \quad .
\end{eqnarray} 
The form of the rotation curves of galaxies is    fixed by the Newtonian potential  via the equation
\begin{equation}
  v(r) =  \sqrt{ r  \frac{d\Phi(r)}{dr} }   \quad ,
\end{equation}
where $v(r)$ is the velocity of an object lying at a distance $r$ from the center. Using the expressions obtained in   Eq(\ref{five})   and the Formula displayed in  Eq(\ref{fourth})  , one obtains in the asymptotic region ( $r \rightarrow \infty $ ) the approximation

\begin{eqnarray}
 \frac{d \Phi^{(1)}(r)}{dr}  &=&   \frac{K_1 }{r} + \frac{K_2}{r^2}   \quad  \quad   {\rm with}   \nonumber \\
K_1  &=& -  \frac{ \omega_0^3 c^2}{2 \beta}   \frac{2}{A_0} +   \frac{1}{4} \beta c^2 \omega_0   A_0     \quad
{\rm and }   \quad  
K_2  =   \frac{ \omega_0^3 c^2}{2 \beta}  \left(  \frac{11}{6} -   h_0   +  \frac{2}{A_0} \right)   \quad . 
\end{eqnarray}
Some  exponential terms, being subdominant in the asymptotic region, have been neglected. The flatness of the rotation curves is due  to the fact that the first term in the last equation is the dominant one at large distances; $K_1$ is the velocity at infinity. This result
was obtained in  \cite{{bana}} . Our derivation simply changes the point  of view, writing everything in terms of the physical radius. One may remark that the  constant  $K_2$  is nothing but  the Schwarzchild mass ; dropping it in the zeroth order approximation does not prevent it from coming back in the next one. This is a  generic feature when one solves perturbatively a differential equation. In the very far region, this term will also be dropped.

We simply want to point out that the approximation obtained so far is actually not reliable.  When the first non vanishing contribution to a perturbative expansion is bigger than  the background, one has to go to higher orders to draw  a  solid conclusion.  To be precise, let us look at the time-time
component of the metric. Stopping at the first order one has
\begin{equation}
g_{00} = - c^2    \left(  1+ \frac{1}{c^2 l^2}  \Phi^{(1)}(r)               \right)   \quad ,  \quad  \Phi^{(1)}(r)  = K_1  \ln{\frac{r}{r_0}}   \quad ,
\end{equation} 
$r_0$  being an arbitrary  constant.
Working with this approximation is legitimate only when
\begin{eqnarray}
 \vert  \Phi^{(1)}(r)   \vert  & << &    c^2 l^2  \quad   {\rm i.e  } \quad   r <<  r_{\star}    =  r_0    \exp{\left( \frac{c^2 l^2}{K_1}\right)  }  
   \quad .
\end{eqnarray}
This simple calculation shows that the approximation obtained so far is valid only at a finite distance from the source. Although one may chose the parameters to fit some rotation curves, the problem is that one is never truly in the  asymptotic region. There is no guarantee that beyond $r_\star$ the  velocities will have the same behavior. We make more comments and remarks in the conclusions.

In the same way
\begin{equation}
g^{-1}_{rr} =    \left(  1- \frac{2}{c^2 r}     \frac{1}{ l^2}    m^{(1)}(r)               \right)   \quad .   
\end{equation} 
is accurate only when the second term is small compared to the first. This leads to the relation
\begin{equation}
\frac{ 2 \omega_0^3 }{ \beta l^2  A_0 }  <<  1   \quad  ,
\end{equation}
where we have neglected terms depending on the radial coordinate  which vanish at infinity. This second constraint is odd since it would imply that the free parameters of the solutions ($A_0,\omega_0$) of the model can not be chosen at will.

\section{Conclusions}
    We have seen that the perturbative approximation used to claim that Born Infeld gravity could explain  flat  galaxy rotation curves
without dark matter is problematic. Firstly, it  breaks down at infinity. This could be anticipated because one needs  a Newtonian potential which scales like a logarithm of the distance in the asymptotic region. Obtaining this kind of behavior at first order with
a zeroth order in which that potential vanishes everywhere(the  mass was taken to be zero)  is clearly not possible. The second point is that it imposes  unnatural conditions  on the free parameters of the solution describing a spherically symmetric  configuration.

Can a higher order approximation solve the problem? This remains to be seen. One would have  to be  careful and in principle expand also the function $\tilde{k} (r)$ in  terms of the new length scale. This was not done in  \cite{{bana}}; the fact  that  there were three equations of motion
for the functions $\Phi^{(1)}(r) , m^{(1)}(r), \tilde{k}(r)$  may be interpreted  as assuming  $\tilde{k}^{(0)}(r)$ to also be vanishing so that
the function computed was actually   $\tilde{k}^{(1)}(r)$. This choice is of course singular and would have to be changed. The  approximation of order $n$ would then be "reliable" to draw conclusions about rotation curves if in the asymptotic region the contribution of order $n+1$ was subdominant when compared to the sum of its predecessors.

 In fact, the situation is even more complicated because the configuration of   Eq(\ref{sixth})  for  the  second  tensor $q_{\mu \nu}$   is nothing but the Schwarzchild solution. This will remain true no matter which order one considers  for the function $\tilde{k}(r)$. It would be surprising that the perturbations affect only the metric field and not its companion. This is a real question which will  have to be addressed.

To finish, let us come back to the fact that  de Sitter space time is an exact solution of the theory in the absence of matter. If it is the only  one, then this model can not   explain  the rotation curves. But one   knows  that in the modified theory of gravity studied by 
  \cite{{turner}}   for example, the de Sitter and the anti de Sitter  spaces are both solutions in vacuum. Something similar may still salvage the claim here.

\end{document}